\documentclass[12pt]{iopart}

\usepackage{graphicx}

\usepackage{iopams}  

\begin{document}

\title{Forward-Backward (F-B) rapidity correlations in a two step scenario}

\author{\underline{J. Dias de Deus} and J.G. Milhano}

\address{Instituto Superior T\'ecnico/CENTRA, Av. Rovisco Pais, P-1049-001 Lisboa, Portugal}



\vspace{0.5cm}

Two-step scenario models  for particle production are based on: 

\begin{enumerate}
\item  creation of extended objects in rapidity (glasma longitudinal colour fields or coloured strings); followed by 
\item local emission of particles.
\end{enumerate}
The first step guarantees the presence of F-B correlations due to fluctuations in the colour/number of sources, while the second step accounts for local effects such as resonances.

The F-B correlation parameter $b$ is defined via
\begin{equation}
\label{eq:def}
\langle n_B\rangle_F = a+b n_F \, , \quad b\equiv D^2_{FB} / D^2_{FF} \, ,
\end{equation}
where $D^2$ is  the variance. In general, correlations are measured in two rapidity windows separated by a rapidity gap so that F-B short range correlations are eliminated. In the two-step scenario models we write \cite{DiasdeDeus:1997di,Braun:2000cc,Brogueira:2006yk,deus2007},
\begin{eqnarray}
\label{eq:d2fb}
D^2_{FB} \equiv \langle n_F n_B \rangle - \langle n_F \rangle \langle n_B\rangle = {\langle n_F \rangle \langle n_B\rangle \over K} \, ,  \\
\label{eq:d2ff}
D^2_{FF} \equiv \langle n^2_F\rangle -\langle  n_F \rangle^2 =  {\langle n_F \rangle^2  \over K} + \langle n_F \rangle \, ,   
\end{eqnarray}
where $1/K$ is the normalized --- e.g., in the  number of elementary collisions ---  long range fluctuation and depends on centrality, energy and rapidity length of the windows. We have assumed, for simplicity, that local emission is of Poisson type.

From (\ref{eq:def}, \ref{eq:d2fb}, \ref{eq:d2ff}) we obtain 
\begin{equation}
\label{eq:b}
b= {\langle n_B\rangle /\langle n_F\rangle \over 1+ K/\langle n_F\rangle } \, .
\end{equation}
It should be noticed that $b$ may be larger than 1, and that a Colour Glass Condensate (CGC) model calculation \cite{Armesto:2007ia} shows a structure similar to (\ref{eq:b}): $b= A [1+B]^{-1}$ (for a discussion on general properties of (\ref{eq:b}) and on the CGC model, see \cite{deus2007}).

A simple way of testing (\ref{eq:b}) is by fixing the backward rapidity window, or $\langle n_B\rangle$, in the region of high particle density and move the forward window along the rapidity axis. We can rewrite (\ref{eq:b}) in the form
\begin{equation}
\label{eq:b2}
b={x\over 1+K'x} \, , 
\end{equation}
where $K' \equiv K/\langle n_B\rangle$ is a constant and $x\equiv \langle n_B\rangle /\langle n_F\rangle$. In (\ref{eq:b2}), one has $1< x < \infty$ with the limiting behaviour:
\begin{equation}
x \to 1\, , \quad b\to {1\over 1+K'} \, ;\qquad  x \to \infty \, , \ b\to \frac{1}{K'} \, .
\end{equation}
The behaviour of (\ref{eq:b2}) is shown in Figure 1 (drawn for $K' =1$).
\begin{figure}[h] 
   \centering
   \includegraphics[angle=0,width=10cm]{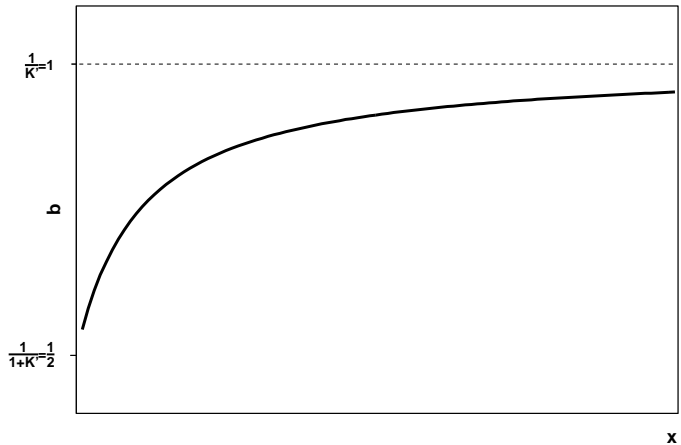} 
   \caption{F-B correlation parameter $b$ (\ref{eq:b2}) with $K' =1$.}
   \label{fig:solution}
\end{figure}

A similar curve is obtained for B-F correlations in the backward region of rapidity.
Note that in $aA$ collisions, $a\leq A$, the centrality and energy dependence of $K'$ is given by \cite{Brogueira:2006yk,DiasdeDeus:2007wb},
\begin{equation}
K'\sim a^{1/2} A^{-1/6} e^{\lambda Y} \, , 
\end{equation}
where $Y$ is the beam rapidity and $\lambda$ a positive parameter. In the symmetric situation, $a=A$ and $K'$ \textbf{increases} with centrality (and the curve of the figure moves down) while in the asymmetric situation, $a=1,2\ll A$ and $K'$ \textbf{decreases} with centrality (and the curve in the figure moves up). As the energy increases $K'$ increases (and the curve moves down).

J.~G.~M.  acknowledges the financial support of  the Funda\c c\~ao para a Ci\^encia e a Tecnologia  of Portugal (contract SFRH/BPD/12112/2003).


\section*{References}

\begin{thebibliography}{6}

\bibitem{DiasdeDeus:1997di}
  J.~Dias de Deus, C.~Pajares and C.~A.~Salgado,
  Phys.\ Lett.\  B {\bf 407} (1997) 335
  [arXiv:hep-ph/9702398].

\bibitem{Braun:2000cc}
  M.~A.~Braun, C.~Pajares and V.~V.~Vechernin,
  Phys.\ Lett.\  B {\bf 493}, 54 (2000)
  [arXiv:hep-ph/0007241].

\bibitem{Brogueira:2006yk}
  P.~Brogueira and J.~Dias de Deus,
  arXiv:hep-ph/0611329.

\bibitem{deus2007}
 J.~Dias de Deus and J.~G.~Milhano,
\textit{in preparation} .
 
\bibitem{Armesto:2007ia}
  N.~Armesto, M.~A.~Braun and C.~Pajares,
  Phys.\ Rev.\  C {\bf 75}, 054902 (2007)
  [arXiv:hep-ph/0702216].
  
\bibitem{DiasdeDeus:2007wb}
  J.~Dias de Deus and J.~G.~Milhano,
  arXiv:hep-ph/0701215.
 

\end{thebibliography}
\end{document}